\documentclass[prb,twocolumn,amsmath,amssymb]{revtex4}
\begin{document}
\author{Kevin Leung}
\affiliation{Sandia National Laboratories, MS 1415, Albuquerque, NM 87185,
{\tt kleung@sandia.gov}}
\date{\today}
\title{The Surface Potential at the Air-water Interface Computed Using
Density Functional Theory}

\input epsf
                                                                                
\begin{abstract}

An accurate prediction of the surface potential ($\phi$) at the air-water
interface is critical to calculating ion hydration free energies and
electrochemical half-cell potentials.  Using Density Functional Theory (DFT),
model interfacial configurations, and a theoretical definition of $\phi$, 
we report a value of +3.63~volt at 0.92~g/cc water density.  A maximally
localized Wannier function analysis confirms that $\phi$ is
dominated by molecular quadrupole (or ``spherical second moment'')
contributions.  We find that the predicted surface potential 
depends on computational details, and conclude that standard DFT codes
and the existing definition of $\phi$ may not yield surface potentials
directly comparable to existing experiments.
 
\end{abstract}

\maketitle


%

\newpage

``Surface potentials'' (or ``potentials of the phase'') are invoked when
a charged particle moves from one phase to another through their mutual
interface.  One of the simplest and most important examples pertains to the
surface between air and
water.\cite{pratt_sur0,guggen,friedman,stillinger,pratt_sur1,pratt_sur2,pratt_sur3,tildesley,dang,brod,hunen1,fawcett,par,gomer,pethica,paluch}
(``Air,'' ''vacuum,'' and ``vapor'' will be used interchangeably in this work.)
The electrostatic potential difference ($\phi$) between
an ion in vacuum at infinity and that ion in pure liquid water contributes a
term $q\phi$ to the absolute hydration free energy ($\Delta G_{\rm hyd}$),
where $q$ is the ionic charge.  $\phi$ thus contributes
to all aqueous media electrochemical half-cell potentials,\cite{truhlar2,sprik1}
which consist of $\Delta G_{\rm hyd}$ plus the pertinent ionization potentials.
Recent Density Functional Theory (DFT) calculations have yielded $\phi$
estimates for the air-water interface\cite{hunt,marsman,ion,cheng,mundy}
substantially larger in magnitude than classical force field
predictions\cite{pratt_sur1,pratt_sur2,pratt_sur3,tildesley,dang} and
experimental values,\cite{fawcett,par,gomer,pethica,paluch} suggesting that the
methodology and physical basis of such calculations should be re-examined.
 
We first focus on molecular simulation
perspectives,\cite{pratt_sur1,pratt_sur2,pratt_sur3,tildesley,dang} 
deferring the question of what is measured in experiments
to the concluding discussions.  For this purpose, we consider a
charge-neutral slab of salt-free liquid water.  Using this kind of
simulation cell, force field-based simulations (i.e., those not based
on electronic structure calculations) have reported
that $\phi$ depends on the water model used,\cite{tildesley,dang,brod}
but that the widely applied SPC/E model\cite{spce} yields
$\phi$=-0.55~volt,\cite{tildesley} similar to related three-point
point-charge water models, including polarizable ones.\cite{dang}  
 
DFT takes into account electronic
structure and molecular polarizabilites, and might be considered 
an improvement over non-polarizable water models.  As will be shown,
however, care must be exercised when interpreting the results
based on the existing theoretical definition of the surface potential
(see below).\cite{pratt_sur1,pratt_sur2,pratt_sur3,tildesley,dang} 
A DFT work that extrapolates the highest occupied molecular orbital
in gas phase ion/water-clusters to infinite cluster size has estimated that
$\phi \sim 4$~volt\cite{hunt} (in our notation).  Using the VASP
code,\cite{vasp} the Perdew-Burke-Ernzerhof (PBE) functional,\cite{pbe}
a bulk liquid simulation cell, and approximations for surface dipoles,
we have deduced that $\phi=+4.05$~volt.\cite{ion} By considering p$K_{\rm a}$
and proton hydration free energies in DFT simulations,
Cheng {\it et al.}\cite{cheng} have postulated that $\phi=+3.5$~volt.
With explicit air-water interfaces, another DFT calculation 
has predicted $\phi=+3.1$~volt.\cite{mundy}  In this work,
we perform a DFT/PBE $\phi$ calculation with air-water interfaces
to compare with previous predictions.  We also use maximally
localized Wannier functions\cite{wannier} to analyze the results in
terms of molecular contributions, which allows a detailed comparison
of DFT and force-field based work.\cite{pratt_sur0,pratt_sur1,pratt_sur2,pratt_sur3,tildesley,dang,brod,hunen1} 
Our work highlights the dependence of $\phi$ on computational details.
 
The theoretical $\phi$ is given by the difference in the average plateau
values between the liquid and vacuum regions in ${\bar{\phi}}(z)$, where
\begin{equation}
\bar{\phi}(z) =\bigg\langle \int \hspace*{-0.08in} \int dx dy
         \hspace*{0.05in} V(x,y,z) /A \bigg\rangle. \label{bar}
\end{equation}
$V({\bf r})$ is the calculated electrostatic potential at point ${\bf r}$,
$A$ is the lateral area of the simulation cell, $z$ is perpendicular
to the interfaces, and the angular brackets denote statistical averaging.
The analytic expressions derived for surface potentials, and formulas for
dealing with long-range electrostatics within periodic boundary conditions
in general, are based on purely coulombic,
$1/r$ potentials.\cite{pratt_sur0,pratt_sur3,hummer,saunders}
Thus 
\begin{equation}
V({\bf r})= \int d{\bf r}' \rho_e({\bf r}') /|{\bf r}-{\bf r}'|
            + \sum_i Z_i/|{\bf R}_i-{\bf r}| . \label{vasp1}
\end{equation}
Here $\rho_e(r)$ is the valence electron density,
${\bf R}_i$ is the position of pseudo-nuclei $i$, and
$Z_i$ is the pseudo-nuclear charge in the pseudopotentials (PP),
with $Z_{\rm O}$=+6$|e|$ and $Z_{\rm  H}$= +$|e|$.  
By default, VASP instead outputs the negative of Eq.~\ref{vasp1} 
after replacing the last term with the entire local PP
$U({\bf R}_i,{\bf r})$.  The short range, non-coulombic
contribution to $U({\bf R}_i,{\bf r})$ should be removed when computing $\phi$.
This ambiguity arises from the use of PP's, and is absent in all-electron
calculations.  
 
\begin{figure}
\centerline{\hbox{\epsfxsize=3.00in \epsfbox{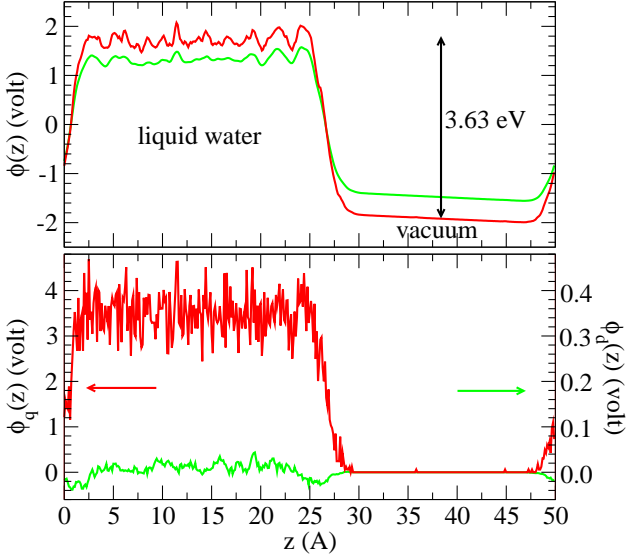}}}
\caption[]
{\label{fig1} \noindent
(a) Electrostatic potential (${\bar \phi}(z)$) 
computed perpendicular to water-vacuum interface
direction $z$, averaged over the lateral ($x$- and $y$-)
directions.  The red and green lines represent predictions
using purely coulomb interactions (Eq.~\ref{vasp1}) and 
$U({\bf R}_i,{\bf r})$ (see text).
$\int_z {\bar \phi}(z) = 0$ over the simulation cell.
(b) Quadrupole (red) and dipole (green) contributions to
${\bar \phi} (z)$.  The latter has its finite electric field removed
and is then expanded ten-fold; the quadrupole component contains more
spatial noise than ${\bar \phi}(z)$.
}
\end{figure}

100 water configurations are taken from a one-nanosecond water-vapor interface
molecular dynamics trajectory generated with 128 SPC/E water molecules in a
50.0$\times$12.5$\times$12.5~\AA$^3$ simulation cell (see the Supporting
Information, SI, for more details).  ${\bar \phi}(z)$ is computed from these
configurations using DFT/PBE and the Eq.~\ref{vasp1} definition.
Figure~\ref{fig1}a shows that $\phi$ averages to
+3.63$\pm0.04$~volt when referenced to vacuum.
If we had used the VASP default and included the short range part of the
PP, $\phi$ would become +2.8~volt instead.
Clearly, the details of the algorithm strongly affect the results.
The vacuum region exhibits a small, 0.005~volt/\AA\, electric field,
indicating that a small but finite net average dipole moment persists in
the water slab due to insufficient statistical sampling.
 
$\phi$ can be rigorously decomposed into
quadrupole and dipole components.\cite{pratt_sur0,pratt_sur3,tildesley,saunders}
In water, they can be written
\begin{eqnarray}
\phi &=& \phi_q + \phi_d  \nonumber \\
 &=& {\bar \phi_q}(z_{\rm water}) -{\bar \phi_q}(z_{\rm air})
        + {\bar \phi_d(z_{\rm water}) }-{\bar \phi_d(z_{\rm air}) };
        \label{phi_qd} \\
{\bar \phi_q}(z) &=& \bigg\langle -(2 \pi /3 A)  \int d{\bf r} \delta(z-R_{m,z})
     \sum_m \rho^m ({\bf r})
        ({\bf r}-{\bf R}_m)^2 \bigg\rangle \, ;  \label{phi_q} \\
{\bar \phi_d}(z) &=& \bigg\langle 4\pi /A \int d{\bf r}
        \sum_m \rho^m({\bf r}) (r_z-R_{m,z}) \Theta(z-R_{m,z})
        \bigg\rangle .  \label{phi_d}
\end{eqnarray}
Here $\rho^m({\bf r})=\rho_e^m({\bf r})+\sum_i Z_i^m \delta ({\bf r}-{\bf R}_i)$
is the charge density of molecule $m$ with all electrons and nuclei $i$
residing on $m$, ${\bf R}_m$ is the molecular center (oxygen atom in
the case of water), and $\Theta$ is the Heaviside function.
Calculating $\phi_d$ requires an interfacial geometry.  As an interface
was not present in Ref.~\onlinecite{ion}, $\phi_d$ was approximated using
the SPC/E water value (+0.21~volt).\cite{tildesley}
 
\begin{figure}
\centerline{\hbox{\epsfxsize=3.00in \epsfbox{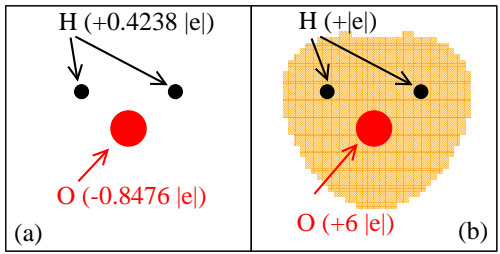}}}
\caption[]
{\label{fig2} \noindent
Charge distributions. (a) SPC/E water model; (b) projection of
DFT/PBE water, using pseudopotentials.  O, H, and electron density exceeding
0.05~$|e|$/\AA$^3$ are depicted in red, black, and orange
respectively.
}
\end{figure}

In contrast, $\phi_q$ can be computed in a bulk liquid simulation cell, with
electron density demarcated into molecular contributions using maximally
localized Wannier functions.\cite{marsman,wannier}  Predicted to be +3.85~volt
at 1.0~g/cc water density,\cite{ion} this VASP/PBE $\phi_q$
is larger in magnitude than and opposite in sign to that for SPC/E water
($\phi_q=-0.76$~volt) because of differences in the charge distributions
(Fig.~\ref{fig2}).  Using the O atom as the molecular center, only the
partial positive charges on the H-sites of SPC/E water contribute to
Eq.~\ref{phi_q}, yielding a negative-definite $\phi_q$.  For VASP/PBE,
Eq.~\ref{phi_q} is instead dominated by the valence electron cloud
surrounding the O-nuclei and $\phi_q$ changes sign.
 
To further analyze the results, we also use maximally localized Wannier
functions to decompose the interfacial ${\bar \phi}(z)$ (Fig.~1a) into
quadrupole and dipole contributions (Fig.~1b).  In the bulk liquid region,
defined as $5 \, {\rm \AA} < z < 20 \, {\rm \AA}$, $\phi_q$ averages to
+3.50$\pm 0.01$~volt.  This differs from the +3.85~volt derived indirectly
(see the SI) because unlike bulk SPC/E water calculations,\cite{spce}
our small lateral simulation cell dimensions dictate a small
Lennard-Jones cut-off distance for SPC/E water, which reduces the density of
the bulk liquid region to 0.92~g/cc.  Equation~\ref{phi_q}
implies that $\phi_q$ is proportional to the liquid water
density.\cite{pratt_sur0,pratt_sur3}  Consistent with this formula,
the 8\% reduction in $\rho_{\rm water}$
and the 9.6~\% decrease in $\phi_q$ compared to the indirect calculation
conducted at 1.0~g/cc water density indeed track each other.
We have not used PBE-based
molecular dynamics to generate water slabs partly because PBE exhibits water
densities that deviate even more strongly from experiments.\cite{siepmann} 
The +3.1~volt DFT value reported previously\cite{mundy} likely also reflects
the low DFT water density present in that work.
 
The small cell also affects $\phi_d$.  After removing the
finite average electric field in the 100 configurations
selected, the SPC/E $\phi_d$ amounts to +0.009~volt, strongly reduced
from the $\phi_d$=+0.21~volt computed in a larger box.\cite{tildesley}
DFT/PBE applied to these SPC/E interfacial configurations
tracks SPC/E results, giving $\phi_d$=+0.012$\pm 0.008$~volt (Fig.~1b).
$\phi_d$ and $\phi_q$ thus add to +3.52~volt, which is consistent
with the +3.63~volt obtained directly using Eq.~1.  The small discrepancy
may arise from the fact that the interfacial simulation cell has prevented
the extrapolation of Wannier estimates of $\phi_q$ to infinite
box size as was done in Ref.~\onlinecite{marsman}.
 
The term ``quadrupole moment'' used to describe $\phi_q$ in Eq.~\ref{phi_q}
is taken from the liquid state
literature;\cite{pratt_sur0,stillinger,pratt_sur3,tildesley}
``radial'' and ``spherical  second moment''\cite{marsman,saunders} have been
used elsewhere. Equation~\ref{phi_q} contains the trace of 
quadrupole tensor, typically set to zero in multiple expansions and does not
contribute to real-space electrostatic interactions.\cite{jackson}
As an example, the atoms in a neon solid are largely spherically symmetric
and exhibit only small multipole (including off-diagonal quadrupole)
moments, but at the equilibrium density of 1.444 g/cc at its melting point,
our DFT/PBE/PP $\phi_q$ estimate for solid neon still amounts to 3.6~volt.

This neon example emphasizes that the theoretical $\phi$ in Eq.~\ref{bar},
computed using the PBE functional, is not a physical quantity.  Indeed, the
PP used omits oxygen core $1s$ electron contributions (see the SI); otherwise
the magnitude
of $\phi$ would be even larger (Eq.~\ref{phi_q}).\cite{note5}
This theoretical $\phi$ is however critical
for DFT-based molecular dynamics calculations of ion hydration free energies
($\Delta G_{\rm hyd}$).\cite{ion,cheng} The reason, familiar in the classical
force field
literature,\cite{pratt_sur0,pratt_sur1,pratt_sur2,pratt_sur3,dang,hunen1}
is that $\Delta G_{\rm hyd}$ at infinite dilution are most conveniently
computed as intrinsic ion
hydration free energies calculated in bulk-water boundary condition simulation
cells using Ewald summation, with the non-ion-specific pure water ``surface
potential'' contribution $q\phi$ added during post-processing.  Ewald technques
arbitarily set the average electrostatic potential to zero over the bulk-water
simulation cell.  The ``true'' (but computation protocol-specific) average
electrostatic potential is restored by referencing the liquid region plateau
${\bar \phi}(z)$ value to vacuum.  This of course involves a rigid shift equal
to $\phi$, which must be obtained using the same definition and pseudopotentials
applied in intrinsic ion hydration calculations.  In DFT calculations, $\phi$
and the intrinsic ion hydration free energy (the latter through Ewald summation
conventions) both contain large, equal but opposite contributions from the
water atomic core regions (Fig.~\ref{fig2}),\cite{marsman} even though ions do
not penetrate into water nuclei.  $\phi$ must be on the order of +4~volt to
yield $\Delta G_{\rm hyd}$ comparable to experimental
data.\cite{ion}  If the VASP/PBE $\phi$ were -0.55~volt
like in SPC/E water, hydration of Cl$^-$ would have been
endothermic and unphysical.\cite{marsman}  Work function calculations in
metals take advantage of a similar cancellation of ambiguities.\cite{lang}
                                                                                
We have so far side-stepped the issue of comparison with experiments.
To the extent that the theoretical $\phi$ is mainly of interest for computing
the absolute hydration free energies of ions at infinite dilution, it can
be treated as a method- or force field-dependent
entity.  Indeed, it has been argued that the Galvani potential
difference between two phases is extremely difficult to measure,
although possible in principle.\cite{pratt_sur0}
$\phi$ also depends non-trivially on the salt present at the
surface.\cite{friedman}  To our knowledge, in the experimental
literature,\cite{fawcett,par,gomer,pethica} there has not been a precise
definition of $\phi$ in terms of microscopic (i.e., electronic and ionic)
properties.\cite{guggen} To the extent that this quantity has been
indirectly measured at the air-water interface, early reported values
strongly varied in magnitude and sign.\cite{par}  Several post-1970 experimental
values are in better agreement with each other,\cite{gomer,fawcett}
yielding +0.025 to 0.16~volt values (see Ref.~\onlinecite{paluch},
Table~1).  They are also in reasonable agreement with the dipole contribution
$\phi_d$ of the SPC/E water model,\cite{tildesley} which raises the intriguing
point that these measurements may predominantly reflect $\phi_d$.
However, these experimental values cannot be directly used to help calculate
DFT-based ion hydration free energies for reasons discussed above.

It may become possible to establish unambiguous experimental
surface potentials in the future.  Here we confine ourselves to the observation
that, just as the theoretical $\phi$ depends on the method used,
the measured $\phi$ may be sensitive to experimental details.
(1) Surface sensitive spectroscopic measurements may yield values for $\phi$
that most closely match all-electron (frozen oxygen $1s$ electrons
or otherwise) DFT results.  They amount to using as probes test particles
that are point charges and do not exhibit many-Fermion effects.
Proposed measurement of $\phi$ using electron
reflectivity\cite{pratt_sur0} with 1-10~keV beams should also probe the
nuclear region, although many-Fermion effects may arise.
(2) If ions are used as experimental probes, e.g., by
considering ion hydration free energies, the solvent atomic nuclear regions
are not sampled.  These regions are mainly responsible for
the large and positive $\phi$ in DFT calculations.  Therefore
a $\phi$ value different from case (1) above should emerge.  It is
conceivable that $\phi$ predicted using classical force fields,
which do not contain distributed charges in atomic nuclear regions,
may be more appropriate here.  (3) In
electrochemical measurements,\cite{fawcett,par,gomer,pethica,paluch} 
electrons are added to/removed from electrodes while ions enter/depart
electric double layers.  Hence a mixture of electrons and ions are
implicitly used as probes.  In light of the present work, the
results should contain $\phi$ contributions from atomic nuclear
regions of electrodes but not from water.  
 
In conclusion, using SPC/E H$_2$O model-derived interfacial configurations,
we have shown that our DFT/PBE pseudopotential calculation directly yields
a +3.63$\pm 0.04$~volt surface potential when we use the theoretical
definition of $\phi$ and a water slab with interior density of 0.92~g/cc.  
Maximally localized Wannier function analysis confirms that this value is
dominated by the density-dependent molecular quadrupole
(or ``spherical second moment'') contribution $\phi_q$, 
consistent with previous calculations.\cite{marsman,ion}  This suggests that
the most robust way to estimate the dominant $\phi_q$ term in DFT calculations
is to use a bulk liquid water simulation cell at 1.0~g/cc density.  The
DFT surface dipole component of $\phi$ is not accurately determined due to
finite size effects, but it tracks the SPC/E value for this simulation cell.
This theoretical $\phi$ is not a physical quantity, in the sense that it is
not what is measured in existing electrochemical
experiments.\cite{fawcett,par,gomer,pethica,paluch} But it is critical for
calculatin ion hydration free energies and modeling electrochemical half cell
reactions.  In general, $\phi$ computed in DFT calculations are sensitive
to simulation details and may not be directly compared to experiments.

We thank Chris Mundy, Shawn Kathmann, Lawrence Pratt, Susan Rempe,
Michiel Sprik, Don Truhlar, and Graham Yelton for interesting discussions,
but stress that not all of them may agree with the perspectives expressed
herein.  We also thank 
Dr. Kathman for sharing Ref.~\onlinecite{mundy} prior to publication.
This work was supported by the Department of Energy under Contract
DE-AC04-94AL85000.  Sandia is a multiprogram
laboratory operated by Sandia Corporation, a Lockheed Martin Company,
for the U.S.~Department of Energy. 

{\bf Supporting Information Available:} 
This material is available free of charge via the Internet at
{\tt http://pubs.acs.org}.


\begin{references}
\bibitem{note5}
The effect of replacing point charges in water models with distributed
charges has also been discussed in Ref.~\onlinecite{pratt_sur2}.
\end{references}
\end{document}